\documentclass[preprint]{elsarticle}
\usepackage{lineno,hyperref}
\usepackage{amsmath}
\usepackage{mathtools}
\usepackage{amssymb}
\usepackage{subfig}
\usepackage{soul}
\usepackage{graphicx}
\usepackage{hyperref}

\newcommand{\bl}{\begin{linenomath}}
\newcommand{\el}{\end{linenomath}}
\newcommand{\be}{\begin{equation}}
\newcommand{\ee}{\end{equation}}

\DeclareMathOperator\SSE{\mathrm{SSE}}

\usepackage{xcolor}
\definecolor{backgreen}{rgb}{0.00, 0.169, 0.212}
\definecolor{textgray}{rgb}{0.514, 0.580, 0.589}

\usepackage{lineno,hyperref}
\usepackage{amsmath}
\modulolinenumbers[1]
\usepackage{verbatim}
\usepackage{amssymb}

\usepackage{xcolor}
\usepackage{mathtools}

\newcommand{\newtext}[1]{\textcolor{blue}{#1}}

\newcommand{\beq}{\begin{linenomath}\begin{equation}}
\newcommand{\eeq}{\end{equation}\end{linenomath}}
\newcommand{\bseq}{\begin{linenomath}\begin{equation*}}
\newcommand{\eseq}{\end{equation*}\end{linenomath}}
\newcommand{\bln}{\begin{linenomath}}
\newcommand{\eln}{\end{linenomath}}
\newcommand{\bal}{\begin{aligned}}
\newcommand{\eal}{\end{aligned}}
\newcommand{\bgat}{\begin{gathered}}
\newcommand{\egat}{\end{gathered}}

\renewcommand{\bar}{\overline}

\journal{Advances in Water Resources}

\begin{document}


\begin{frontmatter}

\title{Entropy: The former trouble with particles (including a new numerical model computational penalty for the Akaike information criterion)}



\author{David A. Benson\corref{mycorrespondingauthor}}
\address{Hydrologic Science and Engineering, Colorado School of Mines, Golden, CO 80401, USA}

\author{Stephen Pankavich and Michael Schmidt}
\address{Department of Applied Mathematics and Statistics, Colorado School of Mines, Golden, CO, 80401, USA}

\author{Guillem Sole-Mari}
\address{Department of Civil and Environmental Engineering, Universitat Polit\`ecnica de Catalunya, Barcelona, Spain}



\begin{abstract}
Traditional random-walk particle-tracking (PT) models of advection and dispersion do not track entropy, because particle masses remain constant.  Newer mass-transfer particle tracking (MTPT) models have the ability to do so because masses of all compounds may change along trajectories.  Additionally, the probability mass functions (PMF) of these MTPT models may be compared to continuous solutions with probability density functions, when a consistent definition of entropy (or similarly, the dilution index) is constructed.  This definition reveals that every numerical model incurs a computational entropy.  Similar to Akaike's \cite{Akaike_1974,Akaike_repub} entropic penalty for larger numbers of adjustable parameters, the computational complexity of a model (e.g., number of nodes) adds to the entropy and, as such, must be penalized.  The MTPT method can use a particle-collision based kernel or an SPH-derived adaptive kernel.  The latter is more representative of a locally well-mixed system (i.e., one in which the dispersion tensor equally represents mixing and solute spreading), while the former better represents the separate processes of mixing versus spreading. We use computational means to demonstrate the viability of each of these methods.
\end{abstract}


\end{frontmatter}

\section{Introduction}
The classical particle-tracking (PT) method was conceived as a means to eliminate numerical dispersion in the simulation of the advection-dispersion equation. Denote a particle position vector in $d$ spatial dimensions by $X$.  The PT method implements an Ito approximation of a Langevin microscopic stochastic differential equation of motion $dX={\bf a} dt + {\bf B} \sqrt{dt}\zeta$, where ${\bf a}$ is a drift vector, $\bf{B}$ is a decomposition of the known diffusion tensor \cite{Lichtner_2002}, and $\zeta$ is a $d$-dimensional vector of independent standard normal random variables.  The probability density function (PDF) of $X$ at some time (denoted here by $c(x,t)$) evolves according to the forward Kolmogorov (or Fokker-Planck) equation
\begin{eqnarray} \label{eq:BKE}
\frac {\partial c} {\partial t} &= & - \nabla \cdot  \left ( {\bf a} c \right ) + \nabla \nabla : \left (\frac{1}{2} {\bf B} {\bf B}^T c \right )
\nonumber \\
& = &  - \nabla \cdot  \left ( {\bf a} c \right ) + \sum_{i=1}^d \sum_{j=1}^d \frac{\partial^2}{\partial x_i \partial x_j}  \left (\frac{1}{2} \sum_{k=1}^d B_{ik}  B_{jk} c \right ).
\end{eqnarray}
So, in order to model conservative transport using the well-known advection-dispersion equation (ADE), we choose specific values of the drift and diffusion terms, namely ${\bf a} = {\bf v} +\nabla \cdot {\bf D}$ and, e.g., ${\bf B} = \sqrt{2 {\bf D}}$
where ${\bf v}$ is a known velocity vector, $\bf{D}$ is the local dispersion tensor that has been diagonalized. The diagonalization of ${\bf D}$ (and the uniqueness of $\sqrt{{\bf D}}$) is justified by the fact that ${\bf D} = \frac{1}{2} {\bf B} {\bf B}^T$ is a symmetric, positive-definite tensor.
In particular, the Langevin equation is then given by $dX=({\bf{v+\nabla \cdot \bf{D}}})dt +\sqrt{2 {\bf D} dt}\zeta$,  (see \cite{labolle, Salamon_2006}), and the resulting density now satisfies the equation
\bl
\be \label{eq:ADRE}
\frac {\partial c} {\partial t}= - \nabla \cdot  \left ( {\bf v} c \right )  + \nabla \cdot \left ( {\bf D} \nabla c \right).
\ee
\el
To approximate the solutions of \eqref{eq:ADRE}, a large number of independent particles are moved according to a forward Euler approximation of the Langevin equation, and the histogram of these particles is used to recreate the density function $c(x,t)$.  Because of the random dispersive motions of particles, the PT method accurately simulates the spread of a plume following the ADE.  But in its raw form, the PT method does not correctly simulate the mixing of dissimilar waters, or dilution of a conservative plume, because particles maintain constant mass.

Mixing only occurs with post-processing of particle positions.  Mixing and/or dilution are commonly measured by borrowing the definition of the entropy $H_D()$ of a discrete random variable $X$ (see the seminal paper by {\em Kitanidis} \cite{Kitanidis_1994} and recent extensions and applications \cite{Chiogna_GRL,Chiogna_WRR,Sund_AWR}).  Entropy is the expectation of the ``information'' contained within the probability density of that random variable.  The information $I(p)$ is a non-negative function of an event's probability $p$ that is defined as additive for independent events, i.e., $I(p_1)+I(p_2)=I(p_1p_2)$.  Because of this axiom, the functional form of information must be $I(p)\propto-\ln(p)$, so that the expected information is also non-negative and defined by
\bl
\be \label{eq:entropy_2}
H_D(X)=\mathbb{E}[I(P(X))]=-\sum_{i=1}^N p(x_i) \ln(p(x_i)),
\ee
\el
for a discrete random variable (RV) with probability mass function $p(x)$ taking non-zero values at points $\{x_1, ..., x_N\}$. By analogy, the continuous analogue of the expected information is
\bl
\be \label{eq:entropy}
H_I(X)=-\int_{f(x)>0} f(x) \ln(f(x)) dx
\ee
\el
for a continuous RV with PDF $f(x)$ [L$^{-1}$].  Because $f(x)$ often will be greater than unity, this definition for a continuous RV may violate the notion of entropy by assuming negative values; therefore, we use the subscript on $H_I$ to represent ``inconsistent'' entropy.  As we show later, this definition is not without its usefulness; however, zero entropy means perfect order (zero mixing) and negative entropy has no physical meaning.  In other words, this definition \eqref{eq:entropy} for a continuous RV is only a loose analogy.  It does not follow from a Riemann-integral representation of \eqref{eq:entropy_2}, meaning
 \bl
\be \label{eq:entropy_not}
\int_{f(x)>0} f(x) \ln(f(x)) dx  \neq \lim_{\Delta x\rightarrow 0}\biggl[ \sum_{i=1}^N f(x_i)\Delta x \ln(f(x_i)\Delta x)\biggr ]
\ee
\el
where $\{x_1, ..., x_N\}$ is a set of values at which $f(x_i) > 0$ for $i=1,...,N$, and the grid spacing $\Delta x = x_{i+1} - x_i$ is uniform for every $i=1,...,N-1$.
In fact, the limit on the right side does not converge for any valid PDF.  In practice, the evaluation of the entropy of some arbitrary continuous function $f(x)$ (like a plume moving through heterogeneous material) that does not have a convenient hand-integrable form, must impose a sampling interval $\Delta V$.  We use this new variable to conform with the usage in \cite{Kitanidis_1994}.  With this finite sampling, an entropy $H_C()$ may be defined that is consistent with $H_D$ in \eqref{eq:entropy_2} by using the approximation that for small $\Delta V$,
\bl\be
\mathbb{P}(x-\Delta V/2 <X<x+\Delta V/2) \approx f(x)\Delta V,
\ee\el
\noindent so that
\begin{eqnarray}
\label{eq:entropyC}
H_C(X)&=&-\int_{f(x)>0} f(x) \ln(f(x)\Delta V) dx
\nonumber \\
&=&-\ln(\Delta V) + H_I.
\end{eqnarray}
Additionally, to construct a discrete approximation of the consistent entropy, we can merely approximate the integral in $H_I$ so that
\begin{equation} \label{eq:entropy_practice}
H_C(X)
\approx -\ln(\Delta V) -\sum_{i=1}^N f(x_i)\Delta x \ln \left (  f(x_i) \right ).
\end{equation}
Now we may identify this sampling volume $\Delta V$ as identical to the volume invoked by {\em Kitanidis} \cite{Kitanidis_1994} to relate the discrete and continuous definitions of entropy, so that $H_D \approx H_C$.  Most commonly, one would let $\Delta V = \Delta x$ in the sum of \eqref{eq:entropy_practice}, but in estimation theory, this discretization may represent different things (Appendix A).  Clearly, the choice of sampling interval $\Delta V$ both allows for a direct comparison of continuous to discrete processes and imposes some restrictions on how entropy is calculated, as we show later.  {\em Kitanidis} \cite{Kitanidis_1994} also defines the dilution index $E$ as the product of the sampling volume and the exponential of the entropy for discrete and continuous random variables.  Using the consistent entropy \eqref{eq:entropy_practice}, this can be written as
\begin{eqnarray} \label{eq:DIC}
E&=&\Delta V e^{H_C}
\nonumber\\
&\approx&\Delta V \exp \bigl[-\ln( \Delta V)- \sum_{i=1}^N f(x_i)\Delta x \ln(f(x_i)) \bigr]
\nonumber\\
&\approx&\exp \bigl[-\sum_{i=1}^N f(x_i)\Delta x \ln(f(x_i)) \bigr].
\end{eqnarray}
As $\Delta x \rightarrow 0$, this uses the classical inconsistent definition of entropy for a continuous random variable, namely $E=\exp[-\int f(x) \ln (f(x))dx] = e^{H_I}$. For a discrete random variable, this becomes
\bl
\be
\label{eq:DID}
E=\Delta V e^{H_D}=\Delta V \exp\biggl(-\sum_{i=1}^N p(x_i)\ln(p(x_i)) \biggr).
\ee
\el
Each definition \eqref{eq:DIC} and \eqref{eq:DID} has units of volume in the number of dimensions of random travel $X$, and has a reasonably well-defined physical meaning as the ``size'' of the volume occupied by either the ensemble of particles or the PDF $f(x)$ \cite{Kitanidis_1994}.

A real or simulated plume of conservative tracer is often idealized as a PDF of travel distance, i.e., the Green's function, when the spatial source is a normalized Dirac-delta function $\delta (x)$.  Without loss of generality, we will only consider plumes that have such a source function, so that we may use concentration as a PDF at any fixed time $T$, so that  $c(x,T)=f(x)$ or some function of $p(x)$ in \eqref{eq:entropy_2} or  \eqref{eq:entropy_practice}, respectively.

The normalized concentration given by the PT method is represented as a collection of the $N$ particles, namely
\begin{eqnarray}\label{eq:conc}
c(x,t)&=&\frac{1}{m_{tot}}\sum_{i=1}^N \int_\Omega m_i\delta (z-X_i(t)) \phi(x-z)dz
\nonumber\\
&=&\frac{1}{m_{tot}}\sum m_i\phi(x-X_i(t)),
\end{eqnarray}
where $c(x,t)$ [L$^{-1}$] is a reconstructed concentration function, $m_{tot}$ is the total mass, $\Omega$ [L] is the physical domain, $m_i$ is the mass of the $i^{th}$ particle, $\delta (x-X_i(t))$ is a Dirac-delta function centered at each particle location $X_i(t)$ for $i=1,...,N$, and $\phi(x)$ [L$^{-1}$] is a kernel function.  The probability of a particle's whereabouts is simply $p(x_i)=m_i/m_{tot}$.  For simplicity here, we will use constant $m_i=m=1/N$, which means that each kernel must integrate to unity and $m_{tot} = 1$. In general, the kernel function is not known or specified in the PT method.  A common choice uses simple binning of arbitrary size $\Delta x$, which is identified with a generalized kernel that depends not merely upon the distance between particle positions and binning grid points, but each separately. In particular, the binning kernel function $\phi(x, X_i(t))$ is defined by
\bl
\be
\phi(x, X_i(t))=  \begin{cases}
    1, & \text{if } x \in [x_\ell, x_{\ell+1}]  \\
    0, & \text{else }
  \end{cases}
\ee
\el
where $\ell = \mathrm{ceil} \left ( \frac{X_i(t) - x_1}{\Delta x} \right )$ is the binning gridpoint to the left of the particle position and $\mathrm{ceil}(x)$ is the ``ceiling'' function.

More recent methods recognize that each particle is a single realization of the Green's function, so that the kernel should have the same shape as $c(x,t)$.  This should be implemented as an iterative process, in which 1) a simple kernel is assumed in \eqref{eq:conc}; 2) an estimated $\hat c(x,t)$ is constructed; 3) a new kernel is estimated $\hat \phi(x) \propto \frac{1}{h} \hat c(\frac{x}{h},t)$ for some $h > 0$, which is then 4) re-used in \eqref{eq:conc} to re-estimate $\hat c(x,t)$ until closure is reached.  The closest approximation of this procedure was given by \cite{Pedretti_kernel}, in which a specific functional form---typically Gaussian---is chosen for $\phi(x)$, and the ``size'' or bandwidth $h$ of the kernel is adjusted based on the centered second moment of the estimated $\hat c(x,t)$.  Because of the convolutional form in \eqref{eq:conc} it is easy to show that the interpolation adds the variance of the kernel to the variance of particle positions, so the bandwidth $h$ of the kernel must be kept small to minimize numerical dispersion.  It is unclear how the ``pre-choice'' of kernel function changes estimates of the entropy, as we discuss in the following section.

\section{Entropy Calculation}
A problem with previous PT methods is that they do not automatically track dilution.  As particles move, they do so as Dirac delta functions (i.e., the kernel itself is a Dirac-delta), and the entropy is based on
\bl\be
c(x,t)=\frac{1}{m_{tot}}\sum_{i=1}^N m_i\delta(x-X_i(t)) = \sum_{i=1}^N\frac1N \delta(x-X_i(t))
\ee\el
so that
\bl
\be \label{eq:H_for delta}
H_D(X)=-\sum_{i=1}^N \frac{ m_i}{m_{tot}} \ln\left(\frac{ m_i}{m_{tot}}\right)= -\sum_{i=1}^N \frac 1N \ln \left (\frac1N \right) = \ln(N).
\ee
\el
Not only does the entropy depend on the number of particles, but it is also constant over all simulation times because $m_i$ and $N$ do not change (although particle-splitting will unnaturally increase entropy).  This also reveals a key feature of particle-tracking algorithms: the use of more particles implies greater entropy (mixing).  This effect was shown in the context of chemical reactions \cite{Benson_react} and measured via concentration autocovariance functions \cite{Paster_JCP}.  


For the particle simulations that follow, we assume a simple problem that is directly solvable: one-dimensional (1-D) diffusion from an initial condition $c(x,t=0)=\delta (x)$.  The solution is Gaussian, with consistent entropy from finite sampling given by:
\begin{eqnarray}\label{eq:entropy_Gauss}
H_C(X) &=& -\int \frac{e^{-x^2/4Dt}}{\sqrt{4 \pi D t}} \ln \biggl (\frac{e^{-x^2/4Dt}}{\sqrt{4 \pi D t}}  \Delta V \biggr) dx
\nonumber\\
&=&-\ln \biggl(\frac{\Delta V}{\sqrt{4 \pi D t}}\biggr)+\frac12
\nonumber\\
&=&-\ln ({\Delta V})+\ln{\sqrt{4 \pi Dt}} + \frac12
\end{eqnarray}
This reveals a few interesting points regarding entropy calculation.  First, for any finite sampling volume, the initial condition has unphysical $H_C=-\infty$. The calculation only makes sense after some ``setting time'' $t>e (\Delta V)^2/(4\pi D) \approx 0.22(\Delta V)^2/D$.  Second, for a reliable estimation of entropy, the sampling interval for a moving plume must remain constant, which means that the sampling volume must be constant in space. For instance, if an Eulerian model possesses finer grids in some areas, the plume will appear to have changing entropy if the Eulerian grid is used for entropy calculation. Third, the sampling interval must be held constant in time.  Very often, PT results are sampled at increasingly larger intervals as a plume spreads out (in order to reduce sampling error, see \cite{Chakraborty}). Clearly, if the sampling size $\Delta V \propto \sqrt{t}$, then the calculated entropy will remain erroneously constant over time.  Fourth, there are two components of the entropy calculation: one given by the PDF, and one given by the act of sampling, or the amount of information used to \emph{estimate} the probabilities.  This implies that, all other things held equal, a finely discretized model has more consistent entropy.  Typically, a model's fitness is penalized by its excess information content, but that is only represented (currently) by adjustable parameters (e.g., \cite{Hill_book}).  The definition of  consistent entropy $H_C$ suggests that the number of nodes or total calculations in a model should also contribute to the penalty.  A simple example and a derivation of a computational information criterion for numerical models is explored in Section \ref{sec:COMIC} and Appendix A.

A general formula that relates entropy growth with the characteristics of the kernel $\phi(x)$ cannot be gained because
\begin{eqnarray}\label{eq:entropy_general}
H(X) &=& -\int \sum_{i=1}^N m \phi(x-x_i) \ln \left (\Delta V m \sum_{i=1}^N \phi(x-x_i) \right) dx
\nonumber\\
&=&-\ln \left ( \Delta V m \right) -m \int \sum_{i=1}^N \phi(x-x_i)\ln \left (\sum_{i=1}^N \phi(x-x_i) dx \right ),
\end{eqnarray}
and the logarithm of the sum inside the last integral does not expand.  As a result, we will rely on numerical applications of several different kernels in computing the consistent entropy \eqref{eq:entropy_practice}.

\section {Mass-Transfer PT Method}\label{sec:MTPT}

A recent PT algorithm \cite{Benson_arbitrary} implements mass-transfer between particles coupled with random-walk particle-tracking (MTPT).  The mass transfer between particle pairs is based on the conceptualization of mixing as a simple chemical reaction (see \cite{Benson_react, Benson_arbitrary}).  Specifically, full mixing between two particles possessing potentially different masses (or moles) $a$ and $b$ of any species $Z$ can be written as the irreversible reaction $aZ+bZ \rightarrow \frac{a+b}{2}Z + \frac{a+b}{2}Z$. This full mixing only occurs between two particles based on their probability of co-location in a time step of size $\Delta t$.   The algorithm has been shown to act as a diffusive operator \cite{Schmidt_accuracy} if the local mixing is modeled as diffusive (i.e., particles move by Brownian motion).  This means that, even if particles are considered Dirac-deltas, their masses continually change, and so the total entropy $H_D$ must also change.  The diffusive nature of the mass transfer may be coupled with random walks to fully flesh out the local hydrodynamic dispersion tensor.  So between diffusive mass transfer, random walks, and local advection, the mass experiences the Green's function of transport (which may be complex due to variable velocities, see e.g., \cite{Benson_Poise}).  A key feature of this algorithm is that the number of particles encodes the degree of interparticle mixing, which is separate, but related to, the spreading of a diffusing plume \cite{Schmidt_accuracy, Benson_Poise}.  Because fewer particles implies greater average separation, the mixing lags behind the spreading of particles to a greater degree as $N$ is decreased \cite{Paster_JCP}.  However, it remains to be shown this effect is shown by the entropy of a conservative plume.

To briefly review, the mass-transfer PT method calculates the probability of collision between particles.  This probability becomes a weight of mass transfer \cite{Benson_arbitrary,Schmidt_accuracy}, with the understanding that co-located particles would be well-mixed.  As a result, for the $i^{th}$ particle, the mass of a given species $m_i$ satisfies
\bl
\be \label{eq:transfer}
m_i(t+\Delta t) = m_i(t) + \sum_{j=1}^N \frac{1}{2}(m_j(t)-m_i(t))P_{ij}
\ee
\el
for $i=1,...,N$. For local Fickian dispersion, each particle pair's collision probability is given by
\bl
\be \label{eq:v(s)}
P_{ij}= (\Delta s / (8\pi \eta D_{ij} \Delta t)^{d/2}) \exp(-r^2/(8\eta D_{ij}\Delta t)),
\ee
\el
where $\Delta s$ is the particle support volume, $D_{ij}$ is the average $D$ between the $i$ and $j$ particles, $r$ is the distance between the $i$ and $j$ particles, and $0<\eta<1$ is the fraction of the isotropic diffusion simulated by interparticle mass transfer. The remainder ($1-\eta$) is performed by random walks.  Here we use the arithmetic average for $D_{ij}$. It should be noted that the $\Delta s$ does not actually change the calculation of mass transfer because the probabilities are normalized, namely
\bl
\be
\sum_{j=1}^N P_{ij}=1, \quad \mathrm{for \ all} \quad i = 1, ..., N.
\ee
\el
The calculated probabilities are normalized in this way because mass must either move to other particles (when $i\neq j$) or stay at the current particle (when $i=j$).  When particle masses are not all the same, and particles are close enough to exchange mass, then the masses must also change, and therefore the entropy $H=-\sum_{i=1}^N m_i \ln(m_i)$ must change.  

As discussed in the Introduction, in the presence of dispersion gradients, the moving particles must be pseudo-advected by the true velocity plus the divergence of dispersion.  The probabilities in \eqref{eq:transfer} should automatically adjust for these gradients because the probability of mass transfer is not given solely by $D$ at the $i^{th}$ particle.  Transfer is automatically lower in the direction of lower $D$, as opposed to the random walk algorithm, which moves a particle with a magnitude given by the value of $D$ at the particle (and hence moves it too far into regions of lower $D$).  Therefore, while the mass transfer algorithm has been shown to be diffusive, it should solve the ADE, rather than the forward Kolmogorov (Fokker-Planck) equation.  However, this effect has yet to be investigated, so we provide evidence in Appendix B.

Several researchers \cite{Dani_kernel,Sole-Mari_2017,Sole2018} have suggested that the kernel representing the probability of particle co-location should actually be a function of total simulation time and/or particle number and local density (through the statistics of the particle distribution), and not merely the time interval over which the particle undergoes some small-scale motions. To summarize, these authors perform smoothing in order to most closely solve \eqref{eq:ADRE}, i.e., the mixing and dispersion are both equally modeled by the diffusion term. Another effect of this operation should be to most closely match the entropy of the (perfectly-mixed) analytic solution of the diffusion equation, so we investigate it here.

Recently, \cite{Sole_Mari_SPH} showed that MTPT can be generalized so that particles can use a Gaussian function (kernel) other than the particle/particle collision probability \eqref{eq:v(s)} for the mass transfer. In doing so, the methodology can be made numerically equivalent to smoothed particle hydrodynamics (SPH) simulations.  The choice of kernel has an effect on simulation accuracy \cite{Sole_Mari_SPH}, which we theorize also changes the entropy, or mixing, within the simulations. Specifically, for the mixing reaction we study here, \cite{Sole_Mari_SPH} rewrites the mass transfer function \eqref{eq:transfer} in the more general form
\bl
\be \label{eq:transfer_SPH}
m_i(t+\Delta t) = m_i(t) + \sum_{j=1}^N \beta_{ij}(m_j(t)-m_i(t))P_{ij},
\ee
\el
where
\bl
\be \label{eq:beta_ij}
\beta_{ij}=\frac{2\eta D_{ij}\Delta t}{h^2},
\ee
\el


\noindent and the expression for $P_{ij}$ \eqref{eq:v(s)} is also modified by the kernel bandwidth choice:
\bl
\be \label{eq:v(s)_kernel}
P_{ij}= (\Delta s / (2\pi h^2)^{d/2}) \exp(-r^2/(2h^2)).
\ee
\el
The kernel bandwidth $h$ depends, at any time, on the global statistics of the particle distribution. For this reason, we call it an {\em adaptive} kernel. More specifically, we set it as the value that minimizes the asymptotic mean integrated squared error (AMISE) of a kernel density estimation. The following expression is valid for a density estimation with a Gaussian kernel and particles carrying identical masses \cite{Silverman1986}:
\bl
\be \label{eq:h}
h_{\mathrm{DE}}=\left( \frac{d}{(2\sqrt{\pi})^dN\int (\nabla^2 f)^2\mathrm{d}x} \right) ^{1/(d+4)},
 \ee
\el
where $f$ is the (usually unknown) true distribution of solute mass. For the present diffusion benchmark problem, $f$ is a  zero-mean Gaussian with variance $2Dt$, so the density estimation kernel is Gaussian with \cite{Sole-Mari_2017}
\bl
\be\label{eq:h_DE}
h_{\mathrm{DE}}=1.06N^{-1/5}\sigma=1.06N^{-1/5}\sqrt{2Dt}.
\ee
\el

In the case of MTPT, however, we do not have a variable density of particles with identical masses, but a constant density of particles with variable masses. As an approximation, we replace the number of particles $N$ in \eqref{eq:h} with the equivalent value for which the average particle density $\rho$ would be equal in the two cases 
\bl
\be \label{eq:rho}
\rho=N\int f^2 \mathrm{d}x,
\ee
\el
which allows us to rewrite expression \eqref{eq:h} as an approximation for MTPT:
\bl
\be \label{eq:h_rho}
h_{\mathrm{SPH}}=\left( \frac{d\int f^2 \mathrm{d}x}{(2\sqrt{\pi})^d\rho\int (\nabla^2 f)^2\mathrm{d}x} \right) ^{1/(d+4)}.
\ee
\el
Once again, because of the simple benchmark problem studied herein, there is a very simple solution for the bandwidth, because the distribution $f$ at any time is a Gaussian with variance $\sigma^2=2Dt$.  Furthermore, if $N$ particles are placed within an interval of length $\Omega$ with average spacing $\Omega/N = 1/\rho$ which doesn't change significantly during a simulation, then the bandwidth reduces to
\bl
\be\label{eq:h_Gauss}
h_{\mathrm{SPH}}=0.82\sigma^{4/5} \rho^{-1/5}\approx 0.82(2Dt)^{2/5}(N/ \Omega )^{-1/5}.
\ee
\el
\noindent We have implemented the adaptive kernels as both the density interpolator $\phi$ of the classical random walk at any time (i.e., a Gaussian kernel with variance $h_{\mathrm{DE}}^2$ in \eqref{eq:conc}) and also in the mass transfer coefficient \eqref{eq:beta_ij} and the probability weighting function \eqref{eq:v(s)_kernel} with bandwidth $h_{\mathrm{SPH}}$ in the mass-transfer algorithm \eqref{eq:transfer_SPH}.

\section{Results and Discussion}

All simulations use $D=10^{-3}$ [L$^2$T$^{-1}$] and are run for $t_{final}=1000$ arbitrary time units.  The spatial domain is arbitrary, but for the MTPT method, we randomly placed particles (with zero initial mass) uniformly on the interval [-5,5], which is approximately $\pm 3.5\sqrt{2Dt_{final}}$.  The MTPT method can represent a Dirac-delta function initial condition by any number of particles.  Here we place one particle at $x=0$ with unit mass.   To enable direct comparison of consistent entropy between all of the methods, we chose equivalent average particle spacing and sampling volume of $\Delta V = \Delta x = 10/N$.  We investigate the calculation of entropy and dilution indices for 1) The PT method using bins of size $\Delta x$; 2) The PT method using constant-size Gaussian interpolation kernels; 3) The PT method using adaptive kernels \eqref{eq:h_DE}; 4) The MTPT method using a collision probability kernel size of $\sqrt{4D\Delta t}$; and 5) The MTPT method using adaptive kernels with size given by \eqref{eq:h_Gauss}.   With the latter two  mass-transfer scenarios, we also let the proportion of diffusion by mass transfer (versus random walks) vary and focus on the two cases of $\eta=1$ and $\eta=0.1$ to see the effect of the collision-based versus SPH-based kernel size.
\subsection{PT versus collision kernel MTPT}
\begin{figure}[bht!]
 \centering
 \includegraphics[width=0.9\textwidth]{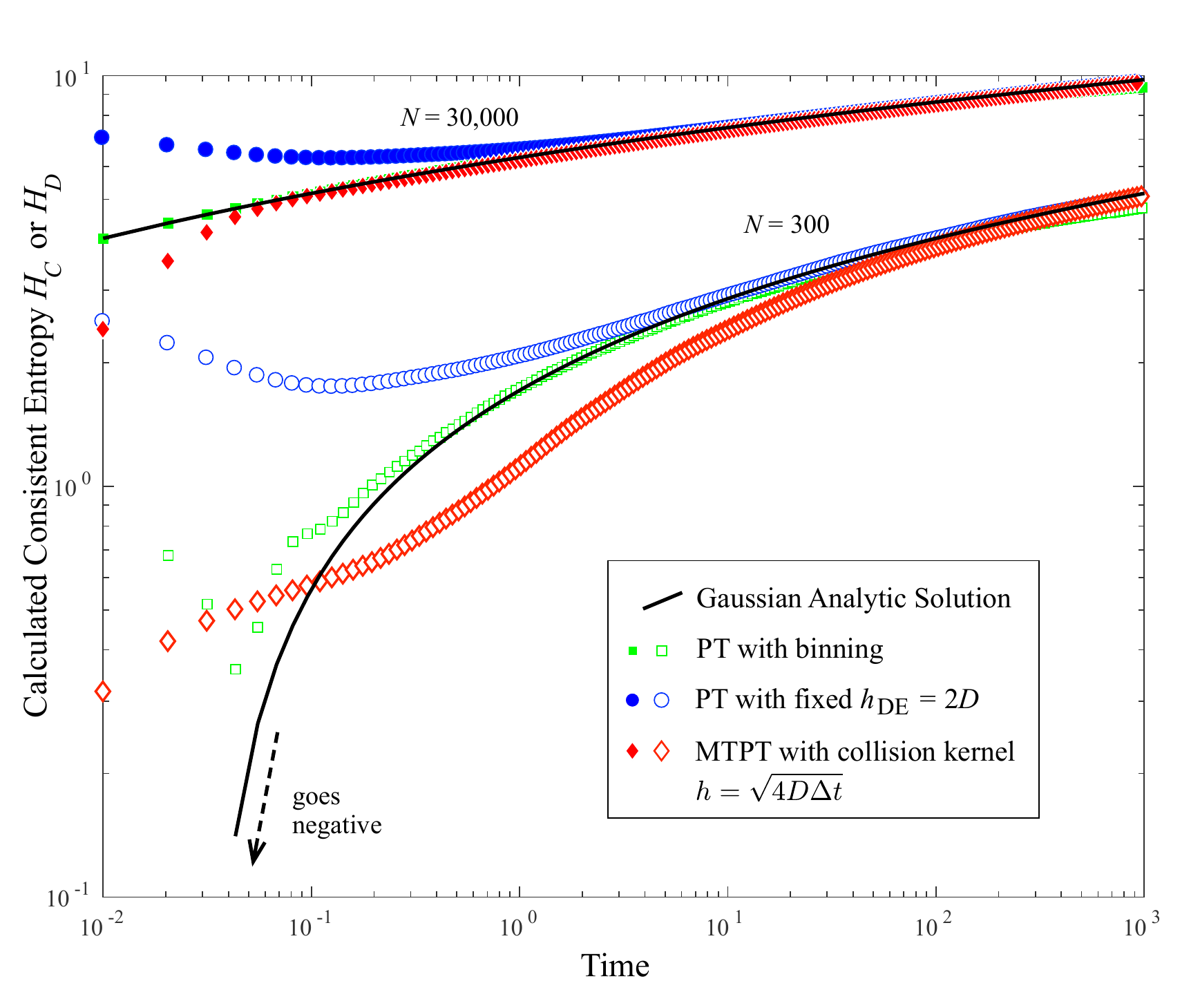}
 \caption{Plot of calculated entropies $H$ and $H_C$ from single realizations of the 1-$d$ random-walk diffusion problem.}
 \label{fig:H_1}
 \end{figure}

\begin{figure}
 \centering
 \includegraphics[width=0.9\textwidth]{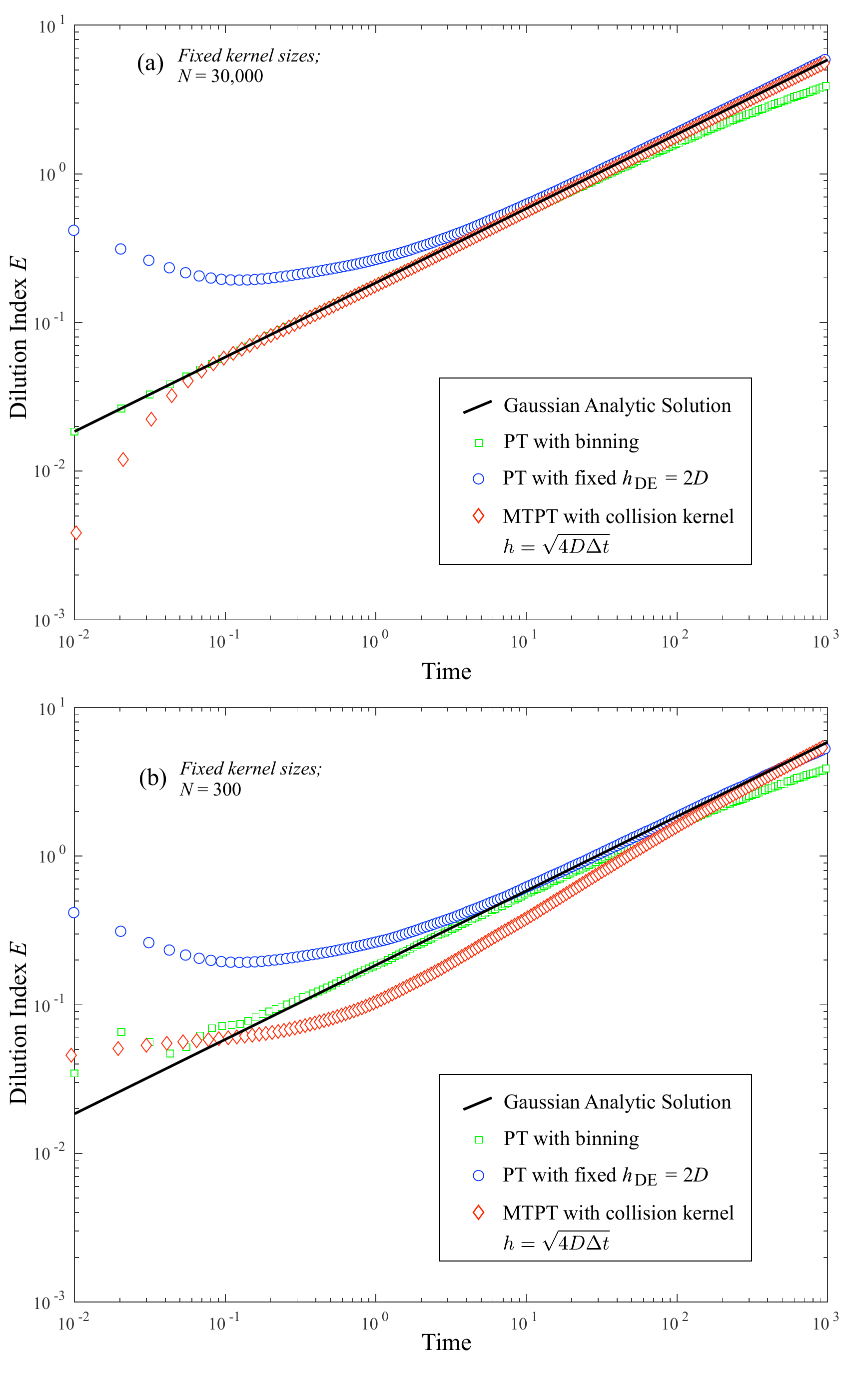}
 \caption{Plots of calculated dilution indices $E$ in the 1-$d$ diffusion problem using interpolation of PT method and MTPT method for ``fixed'' collision kernels: (a) $N=30,000$ and (b) $N=300$.}
 \label{fig:E_1}
 \end{figure}

First, we simulated the classical PT algorithm with concentrations mapped both by binning and by Gaussian kernels with fixed size $2D$.  Because the simulations go from $t=0.01$ to 1000, we chose a kernel size that is too big in the beginning and perhaps too small in the end (i.e., the kernel size is about 1/10 the spread of particles at $t=10$).   The calculated entropies from these simulations were compared to the analytic solution \eqref{eq:entropy_Gauss} and the collision-kernel MTPT algorithm outlined in the previous Section \ref{sec:MTPT}. In these first MTPT simulations, we set the proportion of diffusion by mass transfer $\eta =1$.  In comparison to the other methods, the entropy from binned-PT concentrations matches the analytical solution very well at early times but significantly diverges later (Fig. \ref{fig:H_1}).  The difference between solutions is more obvious when looking at the dilution index $E$ (Fig. \ref{fig:E_1}).  The fixed Gaussian-kernel interpolated concentrations over-estimate entropy and mixing at early time because a fixed kernel size is chosen that is typically larger than the actual diffusion distance for small times.  The MTPT method underestimates entropy at early time relative to the analytic solution \eqref{eq:entropy_Gauss} because the method, by design, does not perfectly mix concentrations.  The random spacings and random walks impart regions where the particles are farther apart, and in these regions, the solutions are imperfectly mixed (i.e., imperfectly diffusive).  As $N$ gets larger, the solution is more perfectly-mixed and converges to the analytic diffusion kernel earlier (Figs. \ref{fig:H_1} and \ref{fig:E_1}).

It is also important to note that neither the analytic solution nor the PT method represent the entropy of the initial condition correctly.  The PT method, with all $N$ particles placed at the origin, still has $H_D=\ln(N)$, while the entropy of the true Dirac-delta initial condition is $H_D=-1\ln(1)=0$.  The analytic solution must use a calculation grid with finite $\Delta x$.  In order for later-time entropies to match, this must be chosen as the same size as the bins for the PT method, i.e., $\Delta x = (x_{max}-x_{min})/N$, where the extents are chosen to almost surely see all particles in a simulation.

On the other hand, the MTPT method can represent the initial condition in many different ways, but here we simply placed one particle at the origin with unit mass, while the remaining $N-1$ particles are placed randomly from the uniform distribution on $-5<x<5$ with zero mass.  Because of this IC, the MTPT method can faithfully represent $H_D{(t=0)}=0$, and the effect of this deterministic, unmixed, IC stays with the simulations for a fair amount of time.  At later time, both the fixed kernel PT and the MTPT methods converge to the analytic solution (Figs. \ref{fig:H_1}, \ref{fig:E_1}).  At early times, however, the fixed kernel interpolator overestimates mixing when generating $c(x,t)$, not only with respect to the Gaussian solution, but also relative to the true initial condition with $H_D=0$.

\subsection{Adaptive kernel versus collision kernel MTPT}
We now turn to simulations using adaptive kernels, in which the particle-particle interaction probability has a time (and particle-number) varying kernel size \eqref{eq:h_rho} in \eqref{eq:v(s)_kernel}.  This is predicated on the fact that a finite sampling of independent random variables is often used to create a histogram of those RVs.  The idea is that a re-creation of the histogram should allow each sample to represent a larger domain than just its value, and a kernel should be assigned to spread each sample value.  In the case of independent, mass-preserving random walks, the idea is clearly sound: for a delta-function initial condition, each particle is a sample of the Green's function, so that each particle's position could be viewed as a rescaled Green's function.  The rescaling depends on the actual Green's function, which may vary in time and space, and the particle numbers.  For independent particles undergoing Brownian motion, the Green's function is Gaussian with variance $2Dt$, and the kernel is shown to be Gaussian with zero mean and standard deviation given by \eqref{eq:h_rho}.  It is less clear that this kernel should represent the particle-particle interaction probability.  First, the global statistics are not important to local reactions, i.e., a paucity of a reactant in one location is not informed by a wealth of reactant outside of the diffusion distance in one timestep.  Second, the masses present on particles are anything but independent, as they depend strongly on their near-neighbors.  Third, the kernels are designed to create a maximally smooth PDF based on random samples, but much research has shown that small-scale fluctuations are the most important driver of reaction rates.  Thus, any kernel that smooths the local fluctuations is artificially increasing reaction rates.  However, much of this discussion is pure speculation, so we implement the kernel functions here as both interpolants of independent random walks and as weights in the reaction function.

\begin{figure}
 \centering
 \includegraphics[width=0.9\textwidth]{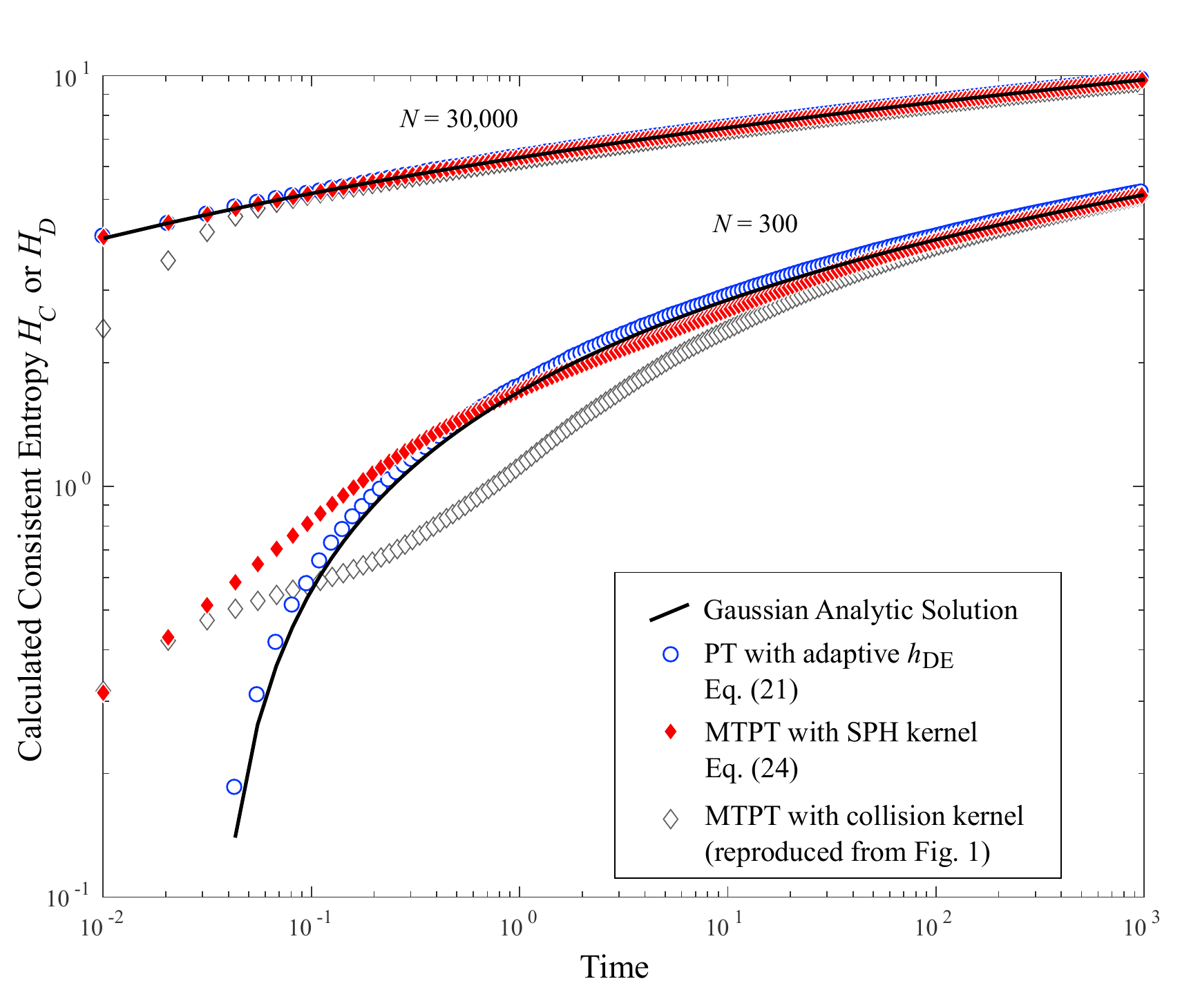}
 \caption{Plot of calculated entropies $H_C$ from ensemble averages of the 1-$d$ random-walk diffusion problem using adaptive kernels for interpolation of simple random walks (blue circles) and for the mass-transfer particle-tracking algorithm (red diamonds) using $N=30,000$ and $N=300$.}
 \label{fig:H_kernel}
 \end{figure}

\begin{figure}
 \centering
 \includegraphics[width=0.9\textwidth]{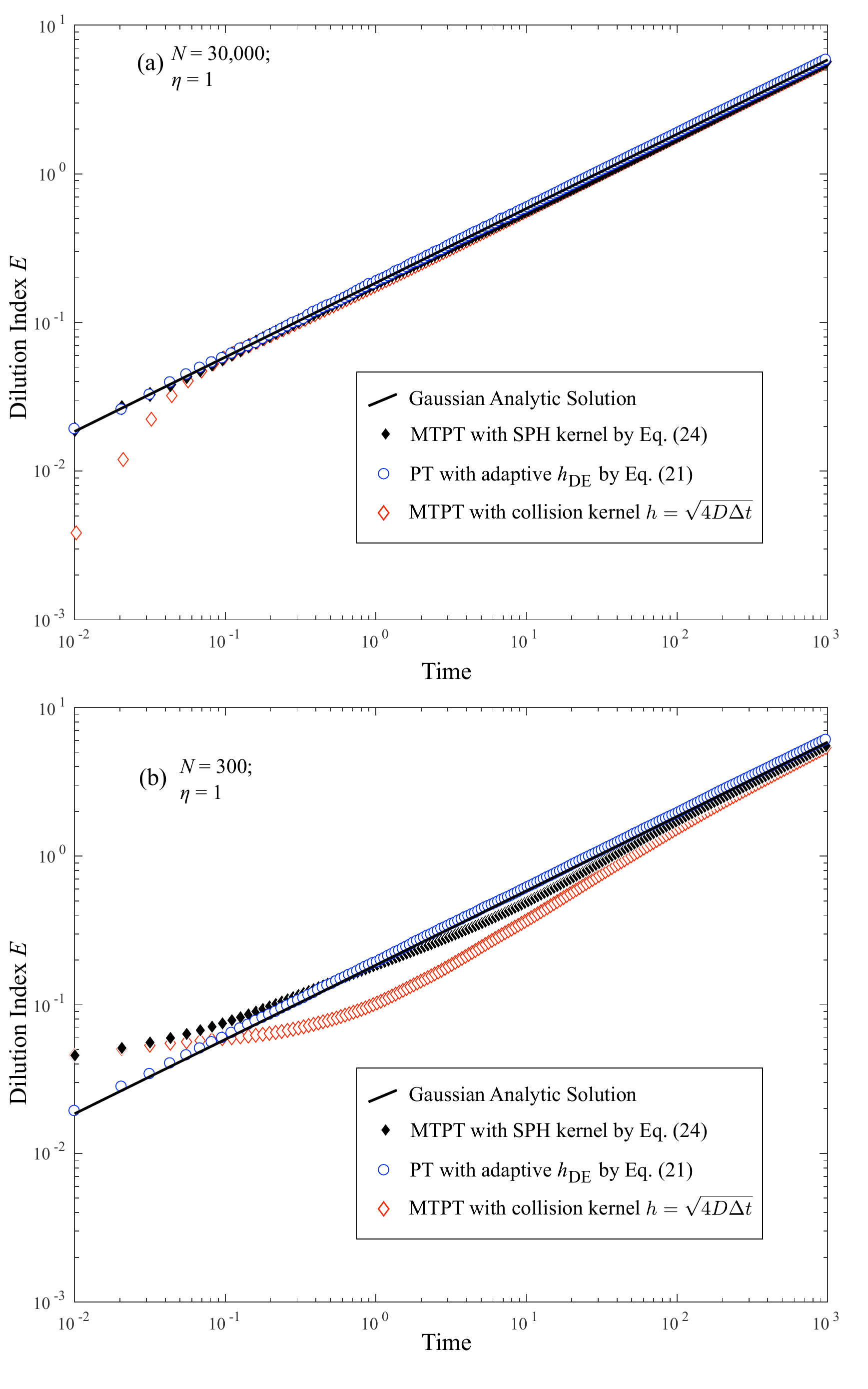}
 \caption{Plots of calculated dilution indices $E$ in the 1-$D$ diffusion problem using adaptive kernels for interpolation of simple random walks (blue circles) and for the mass-transfer particle-tracking algorithm (red diamonds) for (a) $N=30,000$ and (b) $N=300$. MTPT with collision kernel results reproduced from Figure 1 as grey diamonds for comparison.} 
 \label{fig:E_kernel}
 \end{figure}

For brevity and consistency with the previous results, we only show simulations with $N=300$ and $N=30,000$.  Intermediate numbers track the same trends.  For both particle numbers, the kernel-interpolated PT method has consistent entropy and dilution indices that match the diffusion equation analytic solution quite nicely (blue circles, Figures \ref{fig:H_kernel} and \ref{fig:E_kernel}).  The kernels perform exactly as designed for optimally interpolating the PDF of independent, randomly-walking particles.  The adaptive kernels in the MTPT algorithm also match the analytic solution more closely than the collision kernel (black diamonds versus red diamonds, Figures \ref{fig:H_kernel} and \ref{fig:E_kernel}).  The analytic solution assumes perfect mixing, i.e., local mixing and spreading are equal and characterized by the single coefficient $D$.

\subsection{Partitioning of local mixing and random walk spreading}
\begin{figure}[h!]
 \centering
 \includegraphics[width=0.9\textwidth]{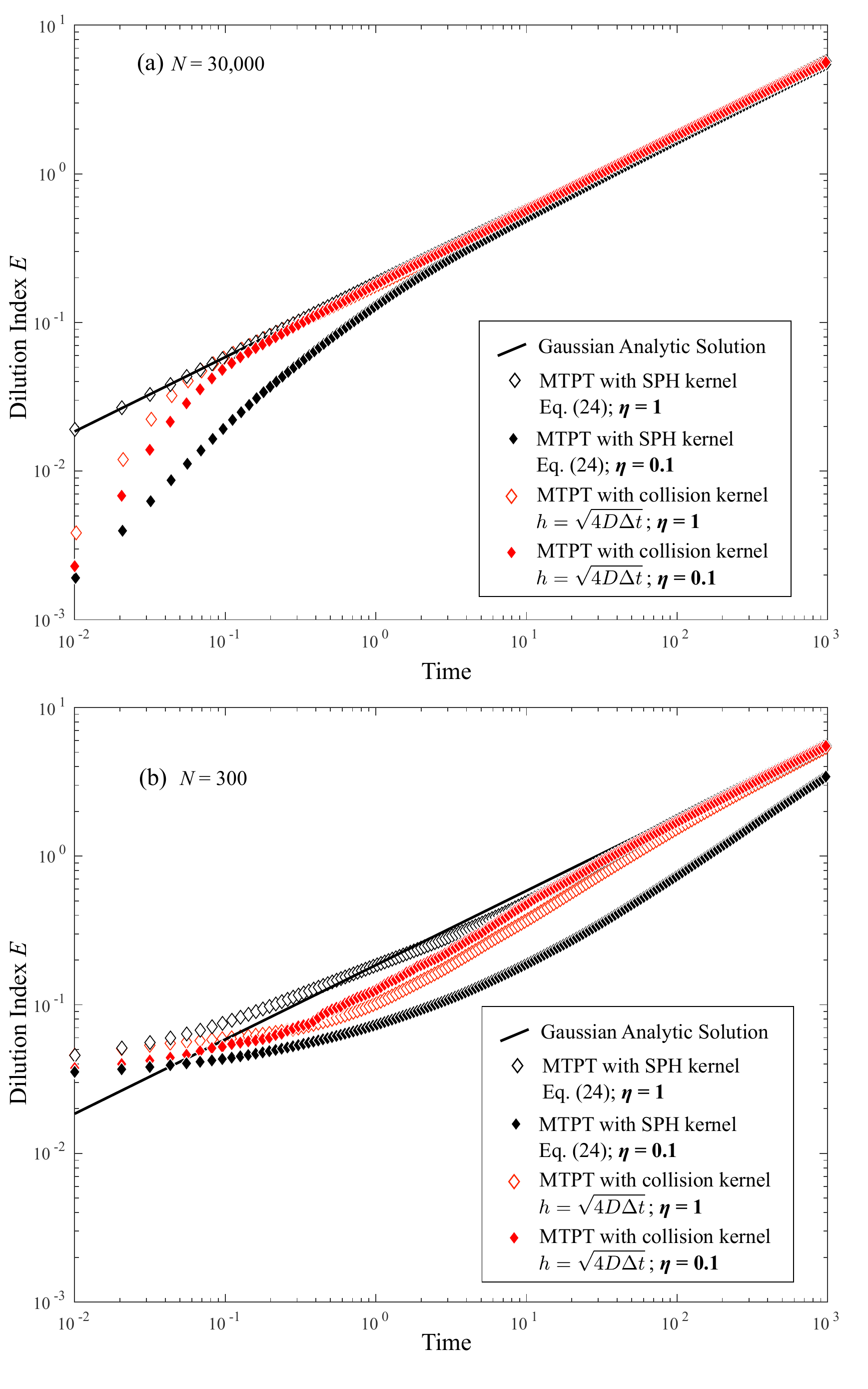}
 \caption{Dilution indices for mixing/spreading proportions $\eta= 1$ and $0.1$ for (a) $N=30,000$ and (b) $N=300$.}
 \label{fig:etas}
 \end{figure}

Recent studies \cite{Schmidt_accuracy,Benson_Poise} that employ the collision kernel for mass transfer have shown that mixing can be simulated as a smaller-scale process than solute spreading. It is unclear whether using the adaptive SPH kernels as defined in \eqref{eq:h_Gauss} can achieve the same effect, given that the particle spreading is part of the evaluation of the kernel size for smaller-scale mixing.  To investigate this effect, we set the mixing proportion $\eta=0.1$ and re-ran the MTPT simulations for $N=300$ and $N=30,000$. Only the dilution indices are shown here, in Figure \ref{fig:etas}. The differences between results for the collision kernel are small, while the adaptive kernel shows significantly decreased mixing.
This increased error for the adaptive kernel when $\eta\ll1$ can be explained as follows. Expression \eqref{eq:h_Gauss} was obtained from \eqref{eq:h_rho} by assuming that the spatial distribution of the solute ($f$) is represented by a Gaussian function with variance $2Dt$. While this is approximately true for $\eta=1$, the micro-scale variability generated when $\eta=0.1$ (see Figure \ref{fig:Gaussians}) suggests that $f$ may not even be continuous and twice-differentiable to start with (which is a requisite for expression \eqref{eq:h_rho} to be valid). Nevertheless, if $\int (\nabla^2 f)^2\mathrm{d}x$ was to be estimated (such as in \cite{Sole2018}), it would be much higher than for a Gaussian $f$ with variance $2Dt$, because of the strong, small-scale concentration variations, suggesting that the truly optimal adaptive kernel obtained from \eqref{eq:h_rho} in this case would be much smaller than \eqref{eq:h_Gauss}.

\subsection{Distributional entropy}
\begin{figure}[h!]
 \centering
 \includegraphics[width=0.99\textwidth]{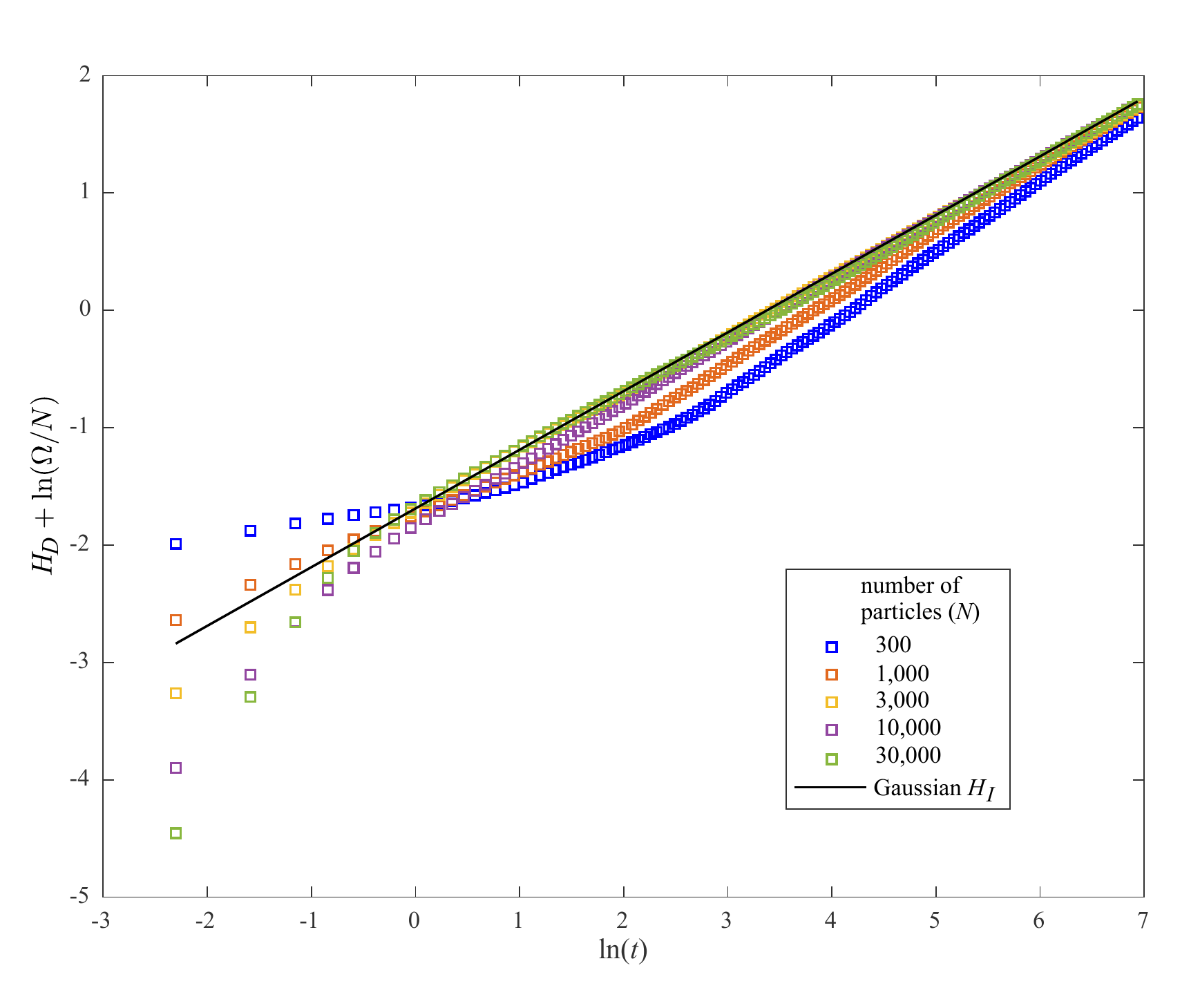}
 \caption{Plots of relative, or PMF, entropy $H_{\text{PMF}}=H_D+\ln(\Omega/N)$ growth over time for different particle numbers diffusing under the MT algorithm. Also plotted is the $H_I(t)$ for a Gaussian diffusion (i.e., eq. \eqref{eq:entropy_Gauss} using $\Delta V=1$).}
 \label{fig:particle_relative_H}
 \vspace{0.2in}
 \end{figure}

As an aside, we note that the particle simulations display greater entropy with more particles. In a similar way that the consistent entropy is related to classically defined entropy for a continuous RV by adding the sampling portion: $H_C=-\ln(\Delta V) +H_I$, the portion of the entropy of a discrete RV can be partitioned into particle number and underlying ``structure'' of the PMF: $H_{\text{PMF}}=\ln(\Omega/N) +H_D $. Using this adjustment, the amount of mixing (given by rate of convergence to the Gaussian) between simulations with different particle numbers can be compared (Fig. \ref{fig:particle_relative_H}).  Here, we ran MTPT simulations using the collision kernel with particle numbers in the set \{100, 300, 1000, 3000, 10000, 30000\}. For smaller $N$, the ensemble average of up to 20 realizations are used because of differences between individual runs.  Quite clearly, the smaller particle numbers have later convergence to the well-mixed Gaussian.  This is a feature of the MT algorithm that is usually reflected in reduced reaction rates.  But a simple measurement of the reduced entropy creation rate with smaller particle numbers is a sufficient demonstration of suppressed mixing.

\begin{figure}[h!]
 \centering
 \includegraphics[width=0.9\textwidth]{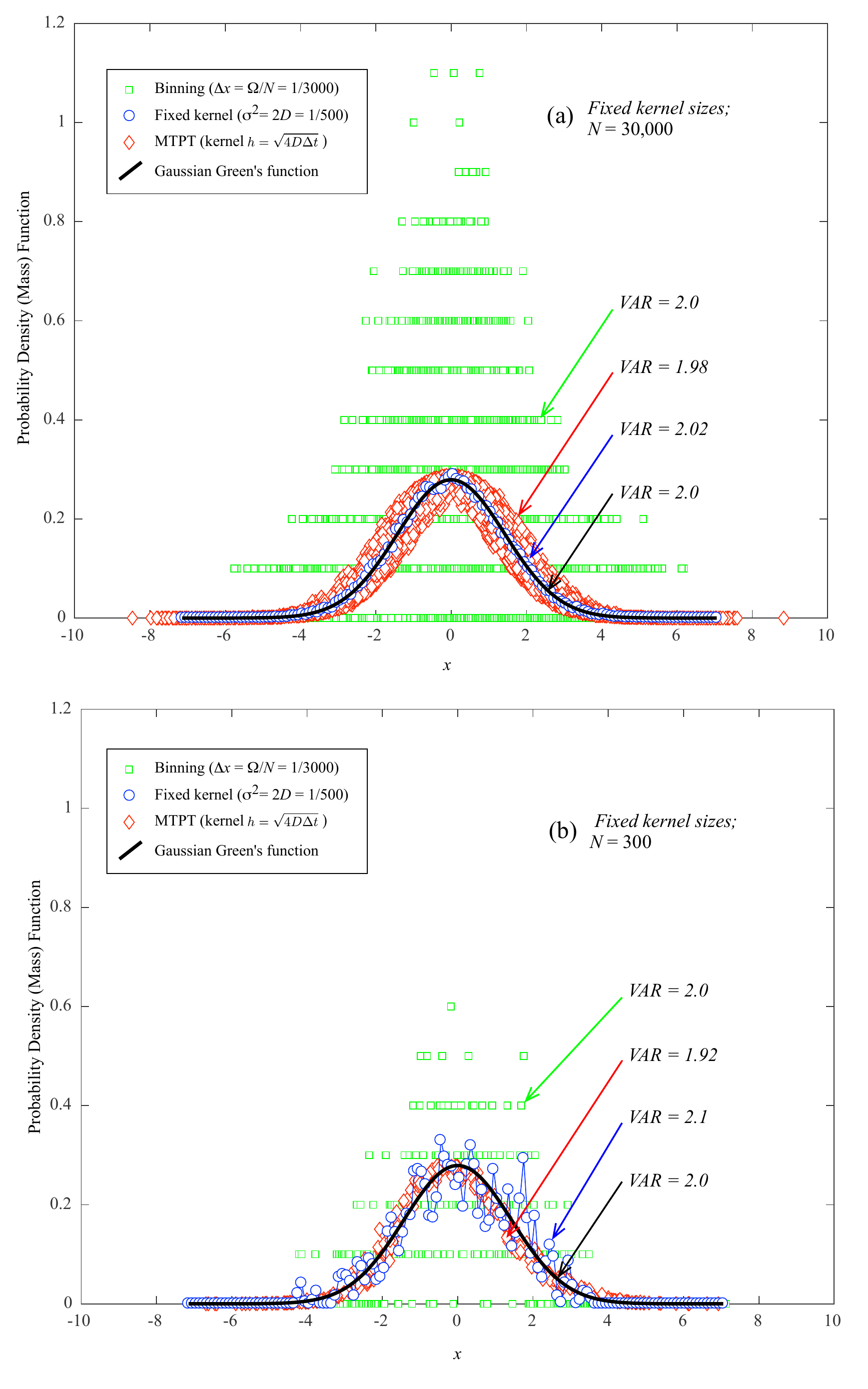}
 \caption{Plot of calculated PMFs and PDFs (and their variances) in the 1-$d$ diffusion problem using ``fixed'' kernels for (a) $N=30,000$ and (b) $N=300$.}
 \label{fig:Gaussians}
 \end{figure}

It is also instructive to inspect the plots of the calculated PMFs and PDFs from the $\eta=0.1$ simulations (Fig. \ref{fig:Gaussians}).  The collision kernel MTPT method is notable because the degree of mixing and the shape of the plume are somewhat independent.  Random walks may place particles with different masses in close proximity, and some time must elapse before local mixing equilibrates those masses.  The result is the mass (or concentration) at any single position in space has substantial variability.  This feature---concentration fluctuations at any point in space---has been exploited to perform accurate upscaling of transport and reaction in heterogeneous velocity fields \cite{Dentz2000,Cirpka2000a,Cirpka2000b,marco_mix_spread,Benson_Poise}.  The fixed kernel interpolation replaces this concentration variance at every location with concentration variability in space.

\section{Computational Entropy Penalty}\label{sec:COMIC}
Philosophically, numerical models provide discrete estimates of dependent variables that may be continuous functions of time and space.  Oftentimes the functions are non-negative and can be normalized to unit area so that they are PDFs.  Therefore, the underlying ``true'' PDF has a certain entropy, and the sampling, or computational, procedure used to approximate these functions adds some artificial entropy because of the information required by the discretization.  One desirable trait of a model is a parsimonious representation of the true physical process, i.e. fewer model parameters are preferred.  At the same time, a more straightforward and accurate computational process is also preferred.  Considerable attention has been paid to parsimonious (few parameter) models, but less attention has been paid to model computational requirements.  Eq. \eqref{eq:entropy_practice} shows that, if a true PDF can be estimated via very few sampling points or nodes, there is less additional entropy incurred in the calculation.  That is to say, if two models (with the same parametric parsimony) yield equivalent estimates of the underlying ``true'' dependent variable, then the model that estimates the PDF with the coarsest sampling, or least computationally intensive structure, is preferred. Replacing the Kulback-Leibler (inconsistent) representation of model entropy with the consistent entropy (Appendix A) gives the COMputational Information Criterion (COMIC) as a natural extension of Akaike's information criterion \cite{Akaike_1974,Akaike_repub}. To emphasize the influence of computational entropy, we illustrate two examples here by estimating a true diffusion given by a Gaussian with variance $2Dt$ by several numerical calculations with zero adjustable parameters (i.e., $D$ is a known parameter).

\subsection{Finite-Difference Example}

\begin{figure}
\vspace{-1.3in}
 \centering
 \includegraphics[width=0.89\textwidth]{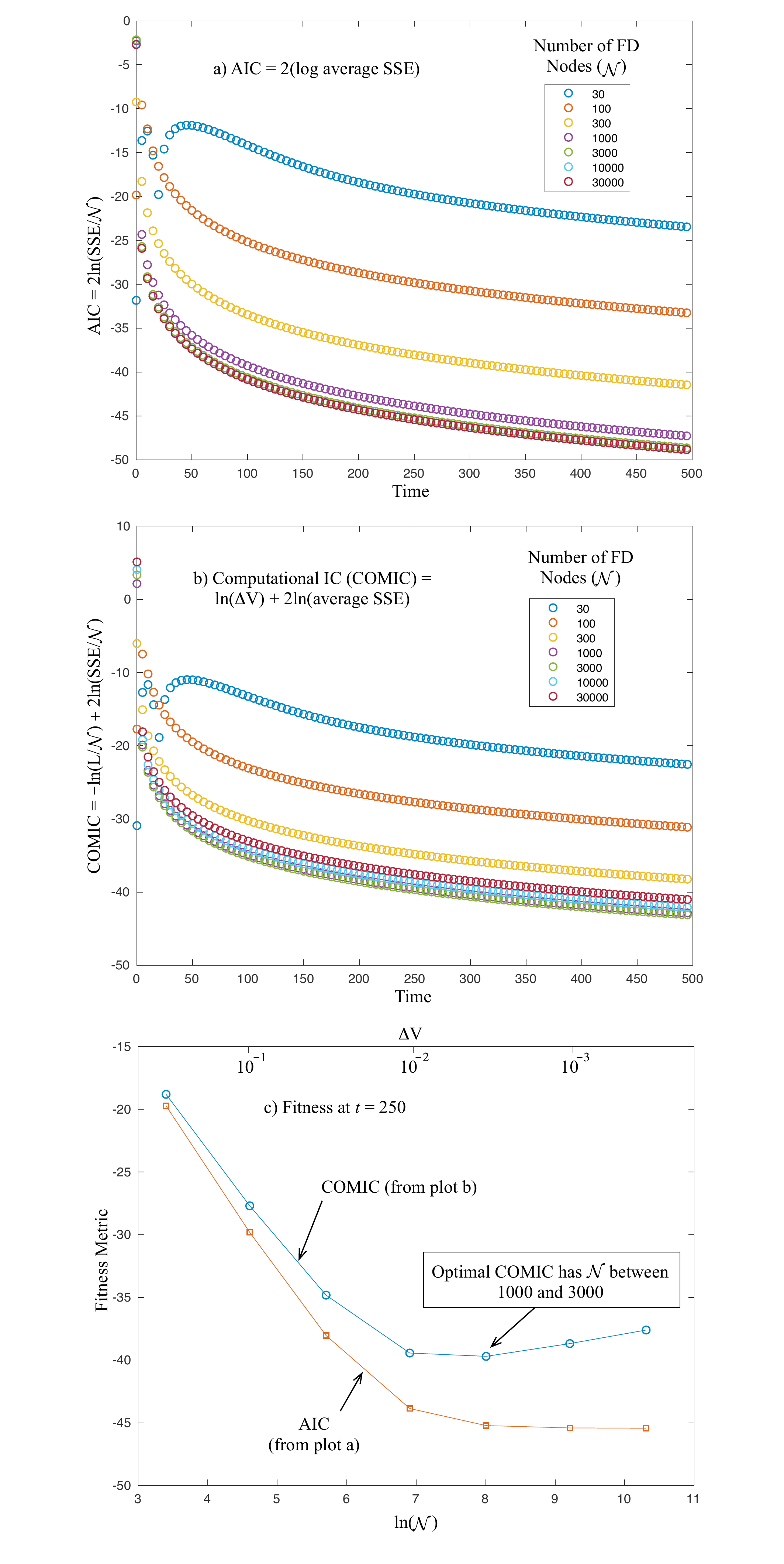}
 \caption{Plots of relative model fitness measures for FD model: (a) log-likelihood function $\ln(\SSE/\mathcal{N})$; (b) computational information criteria COMIC=$-\ln(\Delta x)+\ln(\SSE/\mathcal{N})$; and, (c) both measures versus discretization at a single time $t=250$.}
 \label{fig:FD_fitness}
 \end{figure}

For simplicity, we set $\Delta V = \Delta x = \Omega/\mathcal{N}$ for a fixed domain $\Omega$ and $\mathcal{N}$ nodes, and then compared the numerical estimation of the Green's function of the 1-D diffusion equation given by implicit finite-difference (FD) models with different discretizations $\Delta x \in \{0.4,0.12,0.04,0.012,0.004,0.0012,0.0004\}$.  Other numerical parameters were held constant, including $\Omega=[-6,6]$, $D=10^{-3}$, and $\Delta t =0.05$.  Clearly a smaller $\Delta x$ provides a better estimate of the analytic solution of a Gaussian with variance $2Dt$, but at what cost?  Do 100 nodes suffice?  A million?  Because there are no adjustable parameters, the AIC, which is given by the log-likelihood function ${\text {AIC}}=2\ln(\text{SSE}/\mathcal{N})$, is a decreasing function of the number of nodes $\mathcal{N}$ (Fig.~\ref{fig:FD_fitness}a).  If, however, one factors in the penalty of $\ln(\Delta x)$, there is an optimal tradeoff of accuracy and computational entropy at $\mathcal{N}\approx 3000$ at almost every time step (Fig.~\ref{fig:FD_fitness}b).  Fewer nodes are not sufficiently accurate, and more nodes are superfluous for this particular problem, as shown by plotting the relative fitness criteria (AIC versus COMIC) for each discretization at some time (Fig.~\ref{fig:FD_fitness}c).

Four important points regarding the COMIC immediately arise:
\begin{enumerate}
\item A model is typically sampled at a finite and fixed number of data measurement locations.  We also sampled the many FD models and analytic solution at 15 randomly chosen ``measurement points'' common to all simulations and found nearly identical (albeit more noisy) results.  However, we have not yet investigated the effect of additional sample noise on discerning the optimal discretization.

\item The AIC was derived with the assumption that the number of sample points and computational burden of models is identical and do not contribute to the relative AIC.  Oftentimes the common factors are eliminated from the AIC, and some arbitrary constants are also added, with no effect on {\em relative} AIC.  When looking at the COMIC, however, the choice of likelihood function and inclusion of constants may change the optimal model, so care in the choice of AIC is required.  

\item The numerical solutions are actually conditional densities of the joint densities $c(x,t)$, so that increased number of timesteps should also increase computational entropy (i.e., $\Delta t$ contributes to the multidimensional $\Delta V$, see Appendix A).  Here we held the time step size constant for all FD models, so that the temporal sampling $t = j\Delta t$ has no effect on the {\em relative} entropy.

\item We used a constant spatial discretization $\Delta V = \Delta x$ to simplify the comparative Kullback-Leibler measures.  Some models use variably-spaced grids, so the resulting computational entropy is more complicated than we investigated here.
\end{enumerate}
%

\subsection{Mass-Transfer Particle-Tracking Examples}
\begin{figure}
\vspace{-1.3in}
\centering
\includegraphics[width=0.82\textwidth]{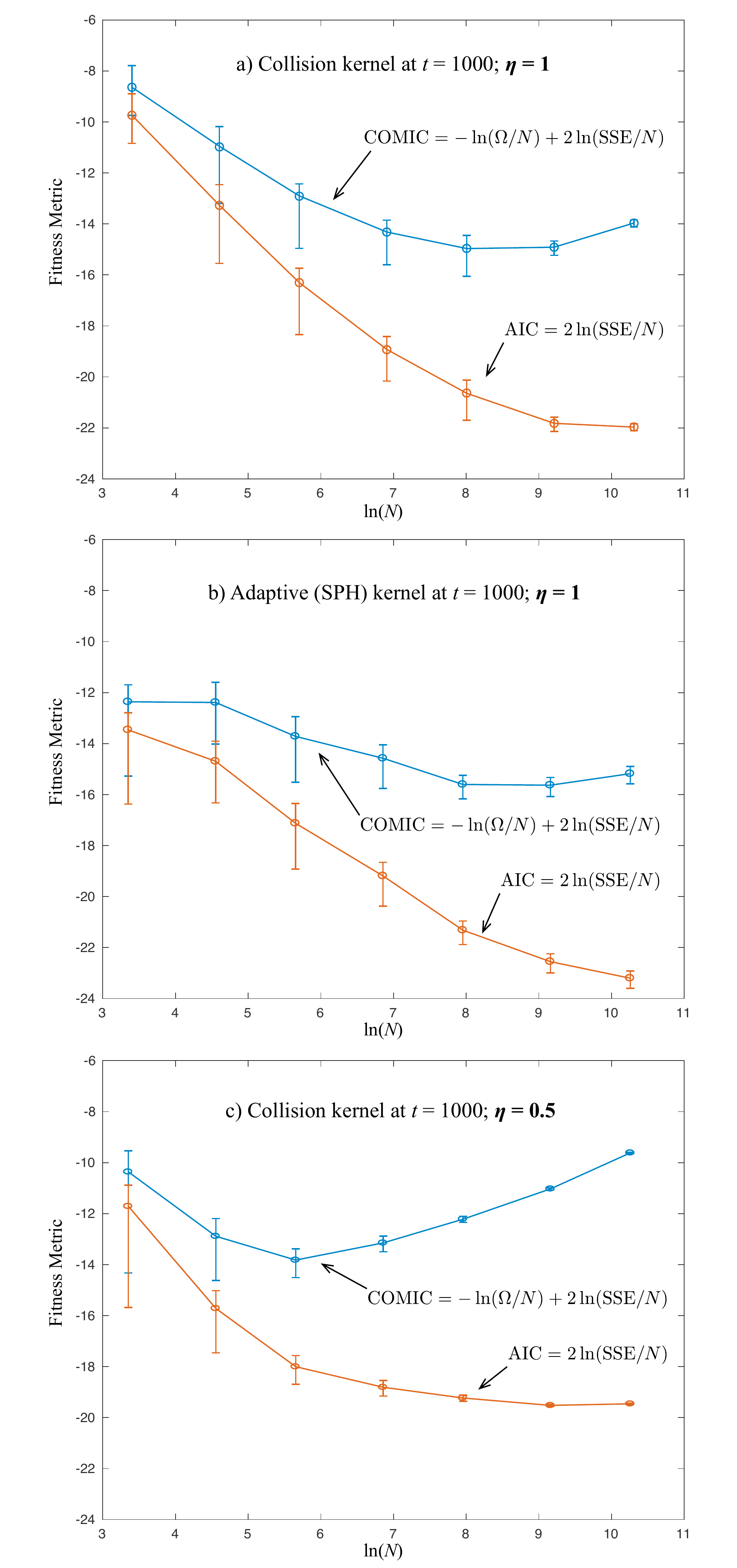}
\caption{Plots of ensemble statistics of relative model fitness measures for three MTPT models of Fickian diffusion at $t=1000$: a) Using the collision kernel with all diffusion by mass transfer ($\eta =1$); b) adaptive SPH kernel using Eq. \eqref{eq:h_Gauss} and full diffusion by mass transfer ($\eta=1$); (b) collision kernel and half diffusion by mass transfer and half by random walks ($\eta=0.5$). Error bars are $\pm$ one standard deviation in ensemble results.}
\label{fig:MTPT_COMICS_2}
\end{figure}

Regarding the last point immediately above for finite-difference models, the main thrust of this paper is the entropy of particle methods. The particles are typically randomly spread in space, so that a constant $\Delta V$ is not possible. However, using the inconsistent entropy isolated the correspondence of the $N$ particles to an underlying PMF (e.g., Fig. \ref{fig:particle_relative_H}). In the case of perfectly-mixed Fickian diffusion, this enables a direct comparison of the fitness of the particle methods to simulating diffusion, and the correction term $-\ln(\Omega/N)$ is the entropy associated with computation. We use this correction, in analogy with the FD results above, to assess the entropic fitness of MTPT methods and test several intuitive hypothesis. First, prior research has shown that fewer particles in the collision kernel MTPT method represent poorer mixing (hence poor fitness when modeling perfectly-mixed Fickian diffusion). In the absence of mixing by random walks (i.e., $\eta=1$), we hypothesize that adding more particles will give smaller, better average SSE, but that the overall model entropic fitness (measured by a smallest COMIC) reaches a maximum at some point. Indeed, a statistically significant minimum is found between $N=1000$ and $N=10,000$ particles, with an estimated minimum at $\approx$ 3,000 particles (Figure \ref{fig:MTPT_COMICS_2}a).


On the other hand, the adaptive SPH kernel is constructed to best match Fickian diffusion, so that the model entropic fitness should be relatively stable across a broad range of particle numbers. This was also found to be true (Fig. \ref{fig:MTPT_COMICS_2}b), and COMIC fitness only suffers in a significant way for $N<100$. Finally, in contrast to the collision kernel for $\eta=1$ (shown in Fig. \ref{fig:MTPT_COMICS_2}a), we hypothesize the splitting the diffusion between mass transfer and random walks will improve the fitness of smaller particle number simulations by eliminating persistent ``mixing gaps'' where large random distances between particles prevents convergence to a well-mixed Gaussian. However, at some point, the model SSE will not improve with the addition of more particles because the ``noise'' of concentrations around the Gaussian will be saturated (see, e.g., Fig \ref{fig:Gaussians}a). Figure \ref{fig:MTPT_COMICS_2}c reveals exactly this behavior in the COMIC: adding random walks decreases the optimal number of particles to $\approx$ 300.

To summarize the MTPT fitness for simulating Fickian diffusion: 1) for the SPH kernel, small particle numbers are sufficient and equally fit (by design); 2) similarly to the FD method, the collision kernel has a minimum COMIC around 3,000 particles; and 3) with the collision kernel, partitioning diffusion by mass transfer and random walks promoted mixing and fitness for smaller particle numbers ($\approx 300$) and clearly shows the superfluous nature of large particle numbers for simulating Fickian diffusion.

 \section{Conclusions}

Classical PT methods do not track entropy until a concentration function is mapped from particle positions.  The choice of bins or kernels for this mapping cannot be arbitrary, as the choice directly changes the entropy, or degree of mixing, of a moving plume. 
The newer mass-transfer method directly simulates entropy without any such mapping (because particle masses continually change), and does so with several beneficial features.  First, the zero-entropy initial condition, and its effect on the early portions of a simulation, are accurately tracked.  Second, the particle number is an integral part of the mixing rate of a plume.  Higher particle numbers simulate more complete mixing at earlier times, as shown by the convergence of entropy to that of a Gaussian.  The MTPT method can use physically-based particle collision probabilities for the mixing kernel, or adaptive kernels dictated by the SPH algorithm.  These adaptive kernels more closely match the analytic Gaussian solution's entropy when solving the diffusion equation in one pass (i.e., all mass transfer given by the diffusion coefficient).   However, when the diffusion/dispersion is split between local inter-particle mixing and spreading by random walks, the adaptive-kernel entropies change substantially and do not match the Gaussian solution for small particle numbers.  The collision kernel does not generate the same effect.  We suggest that the adaptive SPH kernels only be used to solve locally well-mixed problems (i.e., where the dispersion tensor represents both mixing and dispersion equally), whereas the collision kernel may partition mixing and spreading as the physics of the problem dictate \cite{Benson_Poise}.

The fact that discrete (or discretized) approximations to real, continuous functions carry a sampling (or computational) entropy means that metrics which compare different simulations based on information content must be penalized by that computational information.  For this purpose, we define a computational information criterion (COMIC) based on Akaike's AIC that includes this penalty.  We show how a finite-difference solution of the diffusion equation has a well-defined optimal solution of about $3000$ nodes in terms of combined accuracy and computational requirements.  When the MTPT is used to simulate Fickian diffusion, these simulations show that the collision kernel also has a minimum COMIC around 3000 particles, but the SPH kernel, by design, is fit over a large range of particle numbers.  Adding some diffusion by random walks makes the collision kernel a better fit for smaller particle numbers ($\approx N=300$), and shows that simulations of Fickian diffusion for large number of particles is computationally superfluous.  We anticipate that this new entropy-based fitness metric may discount some overly computationally-intensive models that previously have been deemed optimal in terms of data fit alone.

\section{Acknowledgements}We thank the editor and reviewers for extremely helpful comments.  This material is based upon work supported by, or in part by, the US Army Research Office under Contract/Grant number W911NF-18-1-0338. The authors were also supported by the National Science Foundation under awards EAR-1417145, DMS-1614586, EAR-1351625, EAR-1417264, EAR-1446236, and CBET-1705770.  The first author thanks the students in his class ``GEGN 581: Analytic Hydrology'' for inspiring this work.  Two matlab codes for generating all results in this paper (one finite-difference and one particle-tracking) are held in the public repository \url{https://github.com/dbenson5225/Particle_Entropy}. 

\section{Appendix A: Computational Information Criterion and Maximum Likelihood Estimators}
\subsection{Motivation and Theory}

We start with the definition of {\em Akaike's} ``an information criterion'' (AIC) \cite{Akaike_1974}.  The AIC was originally established to select a model and associated parameter values that best fit some given data. In particular, consider a variety of different models defined by distinct parameter vectors $\theta$ and corresponding PDFs $h(y | \theta)$ arising from data values $y_1, ..., y_n$, along with a single vector of ``true'' parameter values $\theta_0$ with PDF $g(y) = h(y | \theta_0)$. The problem of interest is how to optimally select both a number of model parameters $k$ and their associated values $\theta$ to best approximate $\theta_0$ given that we have incomplete knowledge of the latter quantity. In fact, the information provided to make this decision arises only from the given data, which is merely a collection of $n$ independent samples, each representing a realization of a random variable $Y$ with PDF $g(y)$.
Ultimately, the AIC yields an approximate criterion for the selection of parameters, which entails minimizing the quantity
\bl\be
-2\sum_{i=1}^n \ln h(y_i | \hat{\theta}) + 2k.
\ee\el
over the number of parameters $k$, where $\hat{\theta}$ is the maximum likelihood estimate for $\theta$. Furthermore, this process corresponds to maximizing the underlying entropy among such models.

In the context of computing concentrations as in previous sections, we consider a function $c(x,t)$ for which we have a coupled set of data, say $\{(x_i, c_i): i = 1,...,n\},$ 
that represents values of the concentration measured at differing spatial points and at a fixed time $t = T$.
Here, the function $c$ can be a solution to a PDE (e.g., eqn \eqref{eq:ADRE}) or a suitable computational approximation (as in Section $5$), and may depend upon some parameters $\theta$, for instance, $v$ and $D$ in \eqref{eq:ADRE}.
Since the data now has two components, rather than a single variable as in the traditional formulation of the AIC, we first consolidate these into a single vector of values $y_i = (x_i, c_i)$ for $i = 1,...,n$, and consider the joint PDF associated to this data, denoted $h(y| \theta)$. Additionally, we let $\theta_0$ represent the ``true'' parameter values and the underlying PDF be $g(y) = h(y | \theta_0)$.

The selection criterion is based on the entropy maximization principle, which states that the optimal model is obtained by maximizing (over the given data on which $\theta$ depends) the expected value of the log-likelihood function, namely
\bl\be
S(g,h(\cdot|\theta))=\int g(y)\ln(h(y|\theta)) dy.
\ee\el
This quantity is not a well-defined (i.e., strictly positive) counterpart to entropy, as discussed in the main text, and so it is typically implemented in a relative sense among models using the Kullback-Leibler (or relative entropy) measure
\bl
\be \label{KL}
I(g,h(\cdot|\theta))= -\int g(y)\ln \left (\frac{h(y|\theta)}{g(y)} \right ) dy = S(g,g)-S(g,h(\cdot |\theta)),
\ee
\el
which can be interpreted as a measurement of the distance between $g$ and $h$.
As Akaike \cite{Akaike_1974} notes, maximizing the expected log-likelihood above is equivalent to minimizing $I(g,h(\cdot|\theta))$ over the given data.
Of course, since $\theta_0$ is unknown and $g(y) = h(y | \theta_0)$ depends upon knowledge of the ``true'' parameter values, we cannot directly compute $I(g,h(\cdot|\theta))$.
Instead, this quantity must be suitably approximated.
Following \cite{Akaike_1974,Akaike_repub}, if a model $h(\cdot | \theta)$ is close to $g$ and the number of data points $n$ is sufficiently large, a quadratic approximation using the Fisher information matrix can be utilized, and classical estimation techniques imply
\bl\be
I(g,h(\cdot|\theta)) \approx  \biggl (\sum_{i=1}^n \ln g(y_i)- \sum_{i=1}^n \ln h(y_i | \hat{\theta}) \biggr ) + k,
\ee\el
where $k$ is the number of estimated parameters within $\theta$, and $\hat{\theta}$ is the maximum-likelihood estimate for $\theta$. Here, $k$ appears in order to correct for the downward bias introduced by approximating the ``true'' parameter values with their corresponding maximum-likelihood estimates.  Finally, since the first term is constant for any choice of model parameters, it can be omitted in computing the minimization. Therefore, the AIC may be defined (with a scaling factor of two, as in \cite{Akaike_1974}) by
\bl\be
\mathrm{AIC} = -2\ln (\mathrm{maximum \ likelihood})+2k,
\ee\el
\noindent or in the notation described herein
\bl\be
\mathrm{AIC}(\hat{\theta}) = -2\sum_{i=1}^n \ln h(y_i | \hat{\theta}) + 2k.
\ee\el
It is this quantity that one wishes to minimize (over $k$, where $\hat{\theta}$ may depend upon $k$) in order to select the best model approximation to $g$, and this is the basis of our departure.



Though we have not mentioned the process of obtaining the maximum-likelihood estimates $\hat{\theta}$, useful discussions of maximum-likelihood estimators for models with unknown structure are provided in \cite{Hill_book,Brockwell_Davis}. As an example, consider the scenario in which the errors between model and observations are independent Gaussians. In this case the likelihood function is given by
\bl
\be
L(z; \theta)=\left [(2\pi)^n |\Sigma(\theta)| \right]^{-1/2}\exp\left (-\frac{1}{2}z^T\Sigma(\theta)^{-1}z \right),
\ee
\el
where $n$ is the number of observation points, $\Sigma(\theta)$ is a covariance matrix of errors that depends upon some unknown parameter vector $\theta$, and $z$ is a vector of residuals satisfying $z_i = c_i-c(x_i, T)$ for $i=1,..,n$. Recall that $c_i$ is the measured concentration and $c(x_i,T)$ represents the simulated concentration at the spatial data point $x_i$ and time $T$. Therefore, the log-likelihood function is
\bl
\be \label{lnl}
\ln(L)=-\frac{n}{2}\ln(2\pi)-\frac{1}{2}|\Sigma| -\frac{1}{2}z^T\Sigma^{-1}z.
\ee
\el
In practice, the observations are often assumed to be independent, and $\Sigma$ is diagonal. Furthermore, the variance of each observation is often unknown or estimated during the model regression (although numerous approximations can be applied, see \cite{Chakraborty} for assumed concentration errors), so it is assumed that $\Sigma$ depends only upon a single variance parameter, denoted by $\sigma ^2$, and thus satisfies $\Sigma=\sigma^2\mathbb{I}$.  The last term in \eqref{lnl} is more conveniently given in terms of the sum of squared errors $\SSE=z\cdot z = \vert z \vert^2$ (for inter-model comparison), so that
\bl
\be
\ln(L)=-\frac{n}{2}\ln(2\pi)-\frac{n}{2}\ln \sigma^2- \frac{n}{2\sigma^2} \frac{\SSE}{n}.
\ee
\el
Because this function should be maximized, one step in estimation is to take the derivative with respect to $\sigma^2$ and set it to zero, providing an estimator of the observation variance $\sigma^2=\SSE/n$ so that $\ln(L) = -\frac{n}{2} \left (1 + \ln(2\pi) + \ln(\SSE/n) \right ).$  Because the number of observations is usually fixed, the $\frac{n}{2}$ term is canceled from all terms (as maximizing $\ln(L)$ also maximizes $\frac2n\ln (L)$).

Returning to the formulation of the AIC, we encounter a problem with the original derivation applied to the current context, namely we are interested in comparing a discrete model to some true continuous model, and in such a case, it is not proper to compare $g(y)=h(y|\theta_0)$ to $h(y|\theta)$. Rather, we consider $\mathcal{N}$ sampling points denoted by $\{w_1, ..., w_\mathcal{N}\}$, where $w_i = (u_i, v_i)$ is a pair representing the spatial location $u_i$ and computed concentration $v_i$. In a computational model, these $\mathcal{N}$ points merely represent the nodes at which a solution, e.g. a finite difference or particle approximation, is evaluated. Then, as in the definition of entropy, we are comparing the probabilities $g(w_i)dy$ and $h(w_i|\theta)\Delta V$ at each of these $i=1,..,\mathcal{N}$ sampling points where $\Delta V$ is a numerical discretization.
Hence, we can implement the ideas described within the Introduction via \eqref{eq:entropy} and \eqref{eq:entropyC} to construct an analogous discrete approximation to the Kullback-Leibler measure \eqref{KL} that incorporates the sampling volume, namely
\bl
\be \label{KLC}
I_C(g,h(\cdot|\theta))=-\int g(y)\ln \left (\frac{h(y|\theta) \Delta V}{g(y)} \right ) dy = -\ln(\Delta V) + I(g,h(\cdot|\theta)).
\ee
\el
Furthermore, the sampling points need not be the original data points used to select the approximate model. Hence, the natural adjustment analogous to the AIC is based on model computational requirements (or sampling density) given by $\Delta V$ in $d$-dimensions and the number of chosen comparison points. With this, we merely approximate $I(g, h(\cdot | \theta))$ in \eqref{KLC} as Akaike does in order to define an adjusted criterion to the AIC, which we name COMIC or the COMputational Information Criteria, given by
\bl
\be
\mathrm{COMIC}(\hat{\theta}; \Delta V)=-\ln(\Delta V) - 2\sum_{i=1}^\mathcal{N} \ln h(w_i | \hat{\theta}) + 2k.
\ee
\el

In order to focus on the computational implications of this adjustment to the model selection criterion, we consider the case in which the errors between model and observations are Gaussian with variance $\sigma^2$, as in the example illustrated above. In this case, the log-likelihood function evaluated at the maximum-likelihood estimate is proportional to the log of the average sum of squared errors ($\SSE$). Upon removing constants, the form of the COMIC becomes
\bl\be
\mathrm{COMIC}(\Delta V) = - \ln(\Delta V) + 2\ln \left (\frac{\SSE}{\mathcal{N}} \right),
\ee\el
where
\bl\be
 \SSE = \sum_{i=1}^\mathcal{N} (v_i - c(u_i, T))^2
 \ee\el
and $\mathcal{N}$ is the number of comparison point pairs (e.g., data or model nodes), which are denoted by $\{(u_1, v_1), (u_2, v_2), ..., (u_\mathcal{N}, v_\mathcal{N})\}$.  For identical models with equivalent $\SSE$, their measure of distributional entropy is the same, but measurement entropy would be $-\ln(\Delta V)$, so that the model fitness should be adjusted by this measurement, or computational, information.

\section{Appendix B: Effect of $\nabla \cdot {\bf D}$ on the Mass-transfer Algorithm}

We illustrate the effect of spatially-variable ${\bf D}$ in simple 2-$d$ shear flow, borrowing the parabolic velocity profile $v_y=0$ and $v_x=-y^2-by$ of Hagen-Poiseuille flow.  The domain used here is $0<x<400$; $0<y<1$, with concentrations initially zero everywhere except for a strip $90<x<110$ with concentration 1/20, i.e., initial mass=1.  The $x$-domain is periodic, so particles that exit at $x=400$ are re-introduced at $x=0$.  We show a scenario with heterogeneous and anisotropic diffusion ${\bf D} = \bigl[ \begin{smallmatrix}  \alpha_L v_x & 0 \\ 0 & \alpha_T v_x \end{smallmatrix} \bigr]$, with longitudinal and transverse dispersivities $\alpha_L=10^{-2};\ \alpha_T = 10^{-3}$.  Dispersive transport was simulated for $t=500$ with timestep size $\Delta t=1$ either solely by mass transfer or solely by random walks. Because the mass transfer algorithm can move mass among all particles in the domain, a total of 20,000 particles were placed in the $400 \times 1$ domain, with an average of 100 particles in the initial non-zero concentration strip. This gives plenty of ``clean'' particles on either side of the strip.

\begin{figure}[h!]
\vspace{0.5in}
 \centering
 \includegraphics[width=1.0\textwidth]{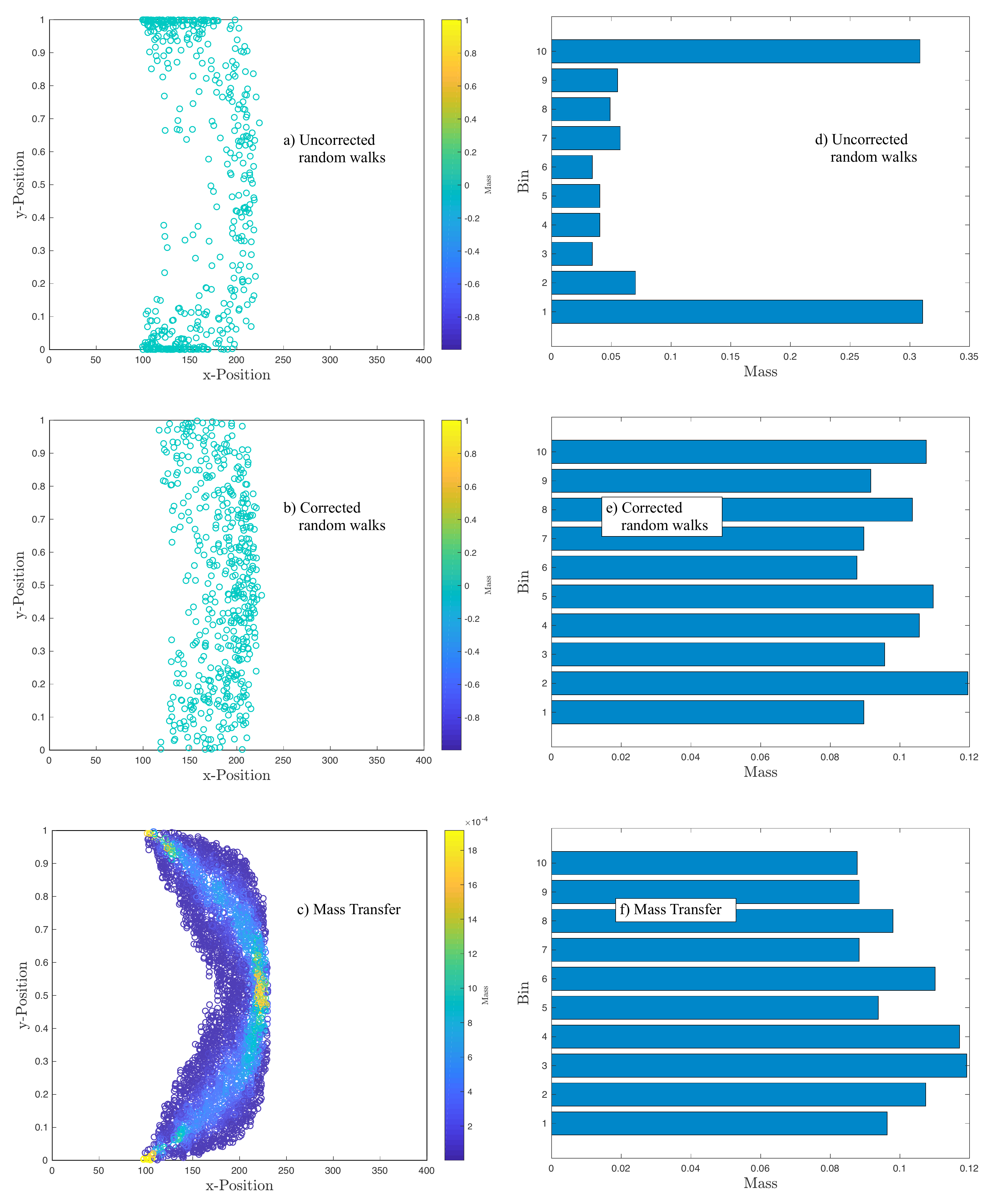}
 \caption{a-c) Particle positions and masses in shear flow simulations. For clarity, only those particles with mass $> 10^{-6}$ are shown. d-f) Histograms on binned masses versus lateral $y$-position. }
 \label{fig:Poise}
 \vspace{0.5in}
 \end{figure}

Pure random walks without the drift correction term migrate all particles, including those with mass, to the lower ${\bf D}$ regions (Figs. \ref{fig:Poise}a).  The drift correction eliminates the lateral bias (Figs. \ref{fig:Poise}b and e).  The mass transfer algorithm has no apparent  bias or need for $\nabla \cdot {\bf D}$ correction (Figs. \ref{fig:Poise}c and f). As an aside, the mass-transfer method quite clearly shows the regions of greatest, and least, shear and mixing (Fig. \ref{fig:Poise}c).

\newpage
REFERENCES
\bibliography{Entropy_4-22-19.bbl}

\end{document}